\documentclass[11pt,twocolumn,prb,superscriptaddress,reprint]{revtex4-1}

\usepackage{graphicx,amsmath,amssymb,color}
\usepackage{tabularx, ctable}
\usepackage{ulem}
 \usepackage{multirow}
\mathchardef\mhyphen="2D

\usepackage{bm}
\usepackage{siunitx}
\usepackage{gensymb}

\usepackage[breaklinks=true,colorlinks,allcolors=blue]{hyperref}

\def\supl{\textcolor{blue}}

\newcommand{\be}{\begin{eqnarray}}
\newcommand{\ee}{\end{eqnarray}}

\begin{document}

\title{Fundamental limits of few-layer NbSe$_2$  microbolometers at terahertz frequencies}

\author{K. Shein$^{*}$}
\affiliation{Moscow Pedagogical State University, Moscow 119991}
\affiliation{National Research University Higher School of Economics, Moscow, 101000.}
\author{E. Zharkova$^{*}$}
\affiliation{Programmable Functional Materials Lab, Center for Neurophysics and Neuromorphic Technologies, Moscow 127495}
\author{M. A. Kashchenko}
\affiliation{Programmable Functional Materials Lab, Center for Neurophysics and Neuromorphic Technologies, Moscow 127495}
\author{A.I. Kolbatova}
\affiliation{Moscow Pedagogical State University, Moscow 119991}
\author{A. Lyubchak}
\affiliation{Moscow Pedagogical State University, Moscow 119991}
\affiliation{National Research University Higher School of Economics, Moscow, 101000.}
\author{L. Elesin}
\affiliation{Programmable Functional Materials Lab, Center for Neurophysics and Neuromorphic Technologies, Moscow 127495}
\affiliation{Institute for Functional Intelligent
Materials, National University of Singapore,  Singapore 117575}
\affiliation{Department of Materials Science and Engineering, National University of Singapore, Singapore 117575}
\author{E. Nguyen}
\affiliation{Programmable Functional Materials Lab, Center for Neurophysics and Neuromorphic Technologies, Moscow 127495}
\author{A. Semenov}
\affiliation{Moscow Pedagogical State University, Moscow 119991}
\author{I. Charaev$^{+}$}
\affiliation{University of Zurich, Zurich, Switzerland 8057}
\author{A. Schilling}
\affiliation{University of Zurich, Zurich, Switzerland 8057}
\author{G.N. Goltsman}
\affiliation{Moscow Pedagogical State University, Moscow 119991}
\affiliation{National Research University Higher School of Economics, Moscow, 101000.}
\author{K.S. Novoselov}
\affiliation{Institute for Functional Intelligent
Materials, National University of Singapore,  Singapore 117575}
\author{I. Gayduchenko$^{+}$}
\affiliation{Moscow Pedagogical State University, Moscow 119991}
\affiliation{National Research University Higher School of Economics, Moscow, 101000.}
\author{D.A. Bandurin$^{+}$}
\affiliation{Department of Materials Science and Engineering, National University of Singapore, Singapore 117575}

\begin{abstract}
The rapid development of infrared spectroscopy, observational astronomy, and scanning near-field microscopy has been enabled by the emergence of sensitive mid- and far-infrared photodetectors. Owing to their exceptional signal-to-noise ratio and fast photoresponse, superconducting hot-electron bolometers (HEBs) have become a critical component in these applications. While superconducting HEBs are traditionally made from sputtered superconducting thin films like Nb or NbN, the potential of layered van der Waals (vdW) superconductors is untapped at THz frequencies. Here, we report the fabrication of superconducting HEBs out of few-layer NbSe$_2$ microwires. By improving the interface between NbSe$_2$ and metal leads connected to a broadband antenna, we overcome the impedance mismatch between this vdW superconductor and the radio frequency (RF) readout circuitry that allowed us to achieve large responsivity THz detection over the range from 0.13 to 2.5 THz with minimum noise equivalent power of 7~pW$\sqrt{Hz}$. Using the heterodyne sub-THz mixing technique, we reveal that NbSe$_2$ superconducting HEBs are relatively fast and feature a characteristic response time in the nanosecond range limited by the slow heat escape to the bath through a SiO$_2$ layer, on which they are assembled, in agreement with energy relaxation model. Our work expands the family of materials for superconducting HEBs technology, reveals NbSe$_2$ as a promising platform, and offers a reliable protocol for the in-lab production of custom bolometers using the vdW assembly technique.

\begin{center}
 $^{+}$ Correspondence to: ilya.charaev@physik.uzh.ch; igorandg@gmail.com;  dab@nus.edu.sg\\
$^{*}$ These authors contributed equally \\
\end{center}
\end{abstract}

\maketitle

Layered van der Waals (vdW) superconductors have recently emerged as a convenient platform to investigate quantum phenomena arising in two dimensions and to prototype future technology~\cite{review_2d,kennes_moire_rev,xing_quantum_grif}. Among them, two-dimensional (2D) niobium diselenide (NbSe$_2$) is one of the most extensively studied compounds~\cite{wang_high-quality,bevolo_specific,frindt_superconductivity,boaknin_heat,staley_electric,monolayer_nbse2}. Not only has it enabled the observation of numerous spectacular effects ranging from high-critical-field Ising superconductivity~\cite{xi_ising} and charged density waves~\cite{xi_strongly,lian_unveiling,CDW_2} to unusual continuous paramagnetic-limited superconducting phase transitions~\cite{sohn_unusual} and magnetochiral anisotropy~\cite{zhang_nonreciprocal,bauriedl_supercurrent}, but it has also given rise to several interesting applications such as superconducting diodes~\cite{bauriedl_supercurrent} and rectennas~\cite{zhang_nonreciprocal}. Despite significant progress in understanding the fundamental properties of this material via transport~\cite{hamill_two-fold} and optical experiments~\cite{hill_comprehensive,jin_multiple,hu_synthesis,sup_photo}, little is known about the response of NbSe$_2$ to terahertz (THz) radiation~\cite{li_high_THz,Nbse2_bol,NbN_spectr}.

The interaction of THz radiation with thin superconductors has been a focal point of intensive research, as superconducting hot-electron bolometers (HEBs) play a crucial role in a variety of applications, ranging from infrared spectroscopy to near-field microscopy and observational astronomy~\cite{klapwijk_engineering,shurakov_superconducting}. Due to the small specific heat capacity of charge carriers, low-dimensional superconductors can convert absorbed low-intensity THz radiation into a strong voltage signal, thanks to the high sensitivity of their resistance to the radiation-induced change in electronic temperature. While conventional superconducting HEBs are typically produced from sputtered Nb or NbN films~\cite{Goltsman_bol}, their thickness and quality are limited by the magnetron sputtering system, making it exceedingly difficult to achieve thicknesses in the nanometer range. Layered van der Waals (vdW) superconducting materials, especially NbSe$_2$, offer an alternative and more accessible platform for this research, owing to the simplicity of achieving few-layer thicknesses~\cite{QualityHeterostructures}. Nevertheless, the opportunity to use NbSe$_2$ for sensitive THz detection remains largely unexplored. The major obstacle for the integration of this material into high-frequency circuitry, critical for the THz exploration in a controllable manner, is the fabrication of low-resistance Ohmic contacts, a known challenge in most transitional metal dichalcogenides. In this article, we overcome this challenge and demonstrate high-responsivity THz detection using antenna-coupled NbSe$_2$ devices, explore the fundamental limits of their application in superconducting HEBs, and provide a recipe for the in-lab production of custom bolometers using the vdW assembly technique.

\begin{figure*}[ht!]
\centering\includegraphics[width=0.98\linewidth]{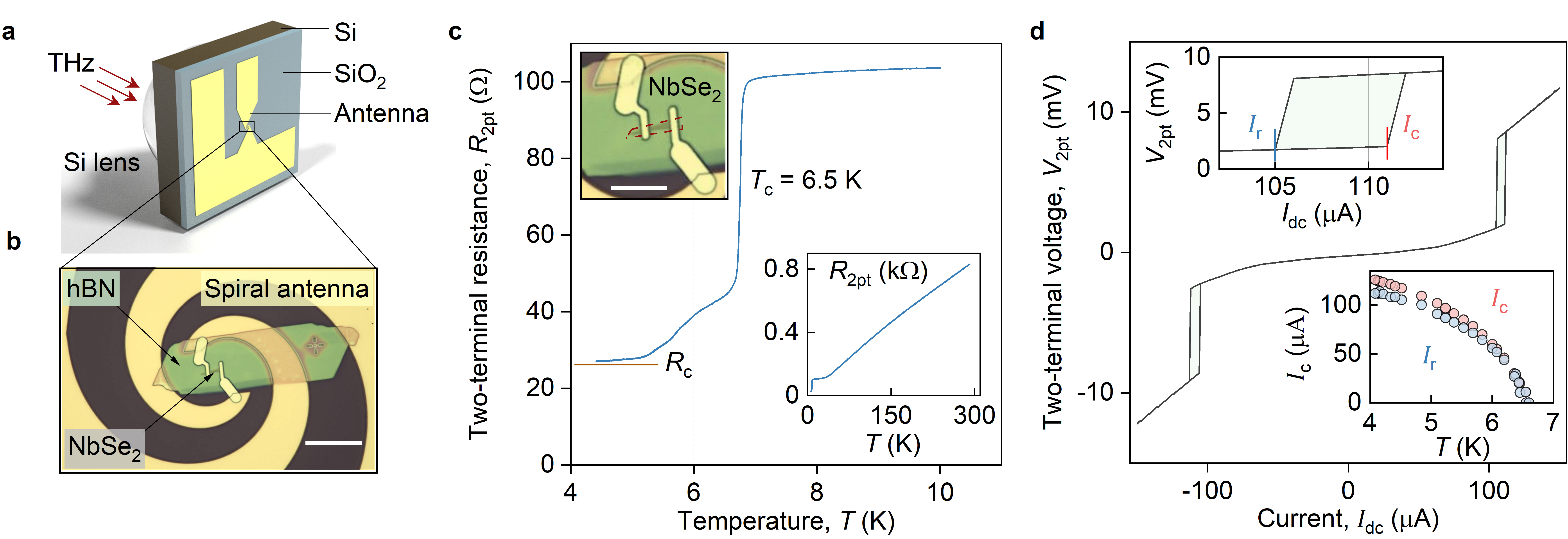}
    \caption{\textbf{Antenna-coupled NbSe$_2$ devices.}
    \textbf{a-b,} Schematic (a) and photograph (white bar is 20 um) (b) of the encapsulated NbSe$_2$ device. Thin NbSe$_2$ microwire is covered by a 50 nm-thick hBN slab and connected by top Al contacts coupled to a log-periodic bow-tie antenna. \textbf{c,} Two-terminal resistance $R_\mathrm{2pt}$ as a function of $T$ for one of our devices close to the critical temperature, $T_c$, and over the whole measurements range (bottom inset). $R_\mathrm{c}$ denotes the residual resistance of the device below $T_\mathrm{c}$. Top inset: zoomed-in photograph of the encapsulated NbSe$_2$ microwire (dashed red line). White bar is 10 um \textbf{d,} Example of the two-terminal $I_\mathrm{dc}-V_\mathrm{2pt}$ curve measured by biasing the microwire with DC current ($I_\mathrm{dc}$). Top inset: $I_\mathrm{dc}-V_\mathrm{2pt}$ curve near the hysteretic region. Bottom inset: The dependence of critical $I_{c}$ and retrapping $I_{r}$ currents on $T$.}
\label{Fig1}
\end{figure*}

Our NbSe$_2$ devices were fabricated using the vdW assembly technique, specifically through the dry transfer method~\cite{Castellanos-Gomez_transfer}. In short, a few-layer NbSe$_2$ microwire-shape flake exfoliated onto a Polydimethylsiloxane (PDMS) stamp was deposited on top of the low-conductivity THz-transparent silicon wafers with a 285 nm thick oxide layer. We intentionally refrained from using monolayer flakes as they have lower critical temperatures $T_\mathrm{c}$ and are prone to degradation upon the contact with environment~\cite{QualityHeterostructures}. The microwire was then covered by a thin slab of hexagonal boron nitride (hBN) to protect it from contamination during electron-beam lithography that was subsequently used to pattern electrical contacts.  Some of our microwires were fully encapsulated by hBN flakes. After lithography patterning, we applied selective reactive ion etching to remove hBN from the contact area \cite{ben_shalom_grapheneJJ}. At the next stage, we applied mild argon milling to strip a thin oxide layer from the surface of the NbSe$_2$ surface and thus prepared the NbSe$_2$ surface for metal deposition (180 nm of Al) conducted in the same vacuum chamber~\cite{sinko_contact}. Such a milling procedure was critical to ensure low interfacial resistance between NbSe$_2$ and the antenna leads essential for the radio frequency (RF) response time measurements (see below). The NbSe$_2$ microwire was further connected to a gold spiral log-periodic antenna (150 nm thick), which was designed to operate over the desired frequency range (see Fig.~\ref{Fig1}a-b). In \supl{Supplementary Information} we provide a detailed description of the fabrication steps along with the parameters of the antenna that we used in this study. In total, we studied four partially-encapsulated or fully-encapsulated devices all exhibiting similar properties. 

Prior to photoresponse measurements, we characterized the transport properties of our NbSe$_2$ devices. Fig.~\ref{Fig1}c shows the dependence of the two-point resistance $R_\mathrm{2pt}$ on temperature, $T$ for one of them. The data for the other samples are shown in \supl{Supplementary Information}. The NbSe$_2$ microwire features a standard metallic behavior as revealed from the monotonic $R_\mathrm{2pt}$ dependence (inset to Fig.~\ref{Fig1}c) and is characterized by a typical to NbSe$_2$ residual-resistance ratio $(RRR)$ of 8~\cite{khestanova_unusual, hamill_two-fold}. The superconducting transition temperatures, $T_\mathrm{c}$, of the microwire, was around 6.5~K which is consistent with previous studies of few-layer NbSe$_2$~\cite{QualityHeterostructures}. Notably, in the middle of the superconducting transition, $R_\mathrm{2pt}\approx50~\Omega$, which is identical to the impedance of the RF readout circuit that ensures the perfect matching between the superconducting flake and the circuitry enabling us to explore the fundamental limits of the response time in NbSe$_2$ superconducting HEBs. At $T<T_\mathrm{c}$, $R_\mathrm{2pt}(T)$ experiences a bulge around 6~K and remains finite $R_\mathrm{c}=27~\Omega$ upon decreasing $T$. While the latter is a measure of the contact resistance between the NbSe$_2$ microwire and the metal lead, the bulge observed in the two-terminal configuration is likely related to the superconducting proximity effect~\cite{proximity_eff}. Figure~\ref{Fig1}d shows an example of the two-terminal $I_\mathrm{dc}-V_\mathrm{2pt}$ curve measured in one of our devices at $T=4.8~$K using DC current ($I_\mathrm{dc}$) biasing and reveals that at this $T$ the critical current, $I_\mathrm{c}$, of the NbSe$_2$ microwire, is of the order of 110 $\mu$A. The finite, yet relatively small slope of the $I_\mathrm{dc}-V_\mathrm{2pt}$ dependence around $I_\mathrm{dc}=0$ is indicative of the aforementioned contact resistance. Also, we observed the hysteresis in the $I_\mathrm{dc}-V_\mathrm{2pt}$ characteristics that we attribute to the metastable state related to the competition between the current-induced Joule self-heating of the microwire in the normal state and electron cooling processes, behavior typical for superconducting micro- and nanowires~\cite{skocpol1974self}. To calculate the exact value of critical current we numerically differentiated the current-voltage characteristics, followed by the finding the maxima of the function. The right inset of Fig.~\ref{Fig1}d shows the $I_\mathrm{c}$ and $I_\mathrm{r}$ vs $T$ dependencies; $I_\mathrm{r}$ is the retrapping current defined in the uppers inset of Fig.~\ref{Fig1}d, which was measured to study the dynamics of thermal relaxation in this material \supl{Supplementary Information}.

\begin{figure*}[ht!]
  \centering\includegraphics[width=0.99\linewidth]{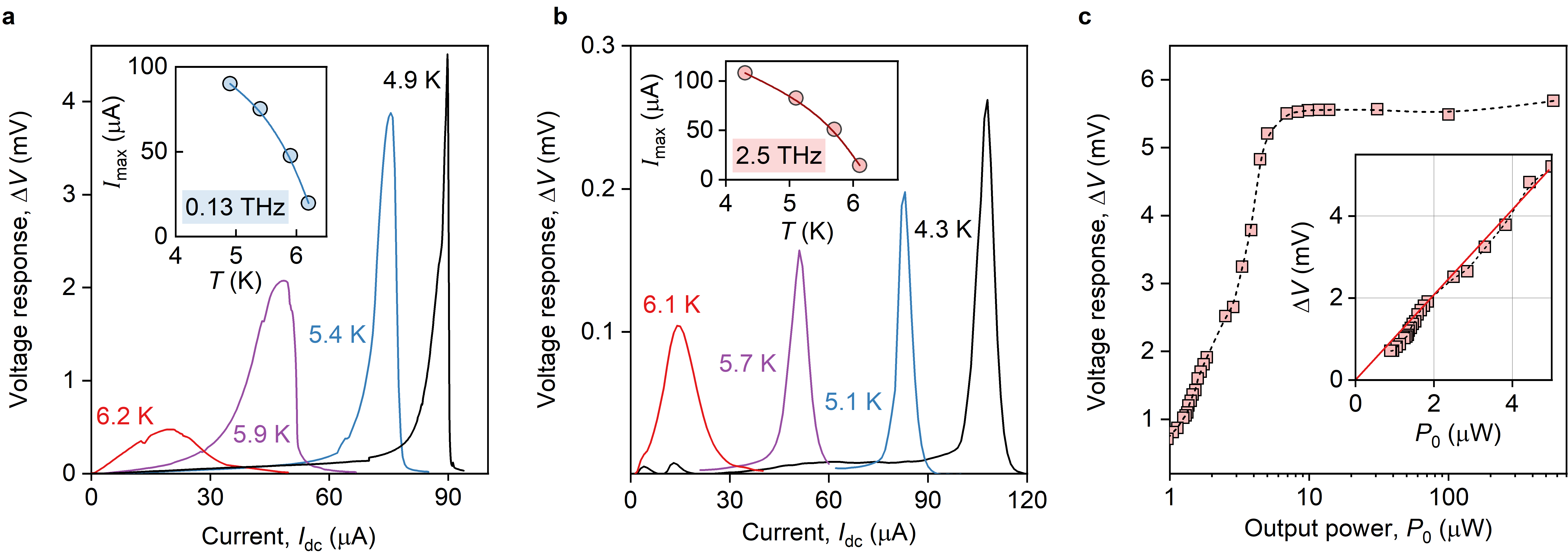}
    \caption{\textbf{THz photoresponse of antenna-coupled NbSe$_2$ microwires.} \textbf{a,} The voltage response $\Delta V$ vs bias current $I_\mathrm{dc}$ at given $T$ obtained under 0.13 THz radiation illumination. Inset: Variation of the current value $I_\mathrm{max}$, at which the $\Delta V$ peak occurs, with $T$. The solid line serves as a visual guide. \textbf{b,} Same as (a), but for 2.5 THz excitation. \textbf{c,} $\Delta V$ as a function of the output power ($P_0$) of the 0.13 THz BWO source. The data is acquired at a temperature of $T=4.8~$K with a bias current $I_\mathrm{dc}=90$~$\mu$A. The dashed line provides a visual guide. Inset: Low-$P_0$ region of the $\Delta V-P_0$ data with a solid red line representing the linear fit.
  }
    \label{Fig2}
\end{figure*}

To perform the photoresponse measurements, we mounted our antenna-coupled NbSe$_2$ microwires onto the hemispherical lens attached to the cold finger of the variable-temperature cryostat equipped with high-frequency coaxial cables. This setting along with the broadband antenna allowed us to carry out direct detection and heterodyne mixing (see below) measurements at different $T$. In the direct detection scheme, we used a backward wave oscillator (BWO) with frequency $f$ = 0.13 THz and a quantum cascade laser generating $f=2.5~$THz radiation. The output radiation was modulated by a mechanical chopper revolving at $77$ Hz while the response voltage $\Delta V$ was measured as a function of bias current $I_\mathrm{dc}$ using a standard lock–in measurement technique synchronized to the chopper frequency~\cite{resterahertz}. The output radiation power, $P_\mathrm{0}$, was determined using a calibrated Erickson power meter or a Golay cell depending on the power range. 

The operation of superconducting HEBs is rooted in the absorption of incident THz radiation, which induces an elevation of the electronic temperature, $T_\mathrm{e}$. Consequently, this results in a significant increase in the sample's resistance when it is brought close to the superconducting transition~\cite{Floet_hot,Gerhenzon_electron_heat,Semenov_heb_rev,Goltsman_bol,Prober_mixer,2.5Thz} . Figures~\ref{Fig2}a-b illustrate the dependencies of $\Delta V$ on the bias current $I_\mathrm{dc}$ when the sample is illuminated with $0.13~$THz and $2.5~$THz radiation. For both  $f$, we observed a pronounced peak in $\Delta V$ near the critical current for a given $T$. As the $T$ is reduced below $T_\mathrm{c}$, the peak's magnitude gradually increases, concurrent with an increase in the current value ($I_\mathrm{max}$) at which it appears (see insets in Figs.~\ref{Fig2}a,b). Above $T_\mathrm{c}$, the peak disappears, naturally indicating its association with the bolometric response, wherein $\Delta V=I_\mathrm{dc}\Delta R$, with $\Delta R$ denoting the radiation-induced change in resistance.

For further characterization, we have measured the photoresponse of our devices upon varying $P_\mathrm{0}$ of the 0.13~THz source using an attenuator coupled to the BWO output waveguide, as the latter provides calibrated power variation (Fig.~\ref{Fig2}c). At low $P_\mathrm{0}<5~\mu$W, $\Delta V$ featured linear scaling with $P_\mathrm{0}$ that allowed us to estimate the voltage responsivity $R_{V}$ =  $ \frac{\Delta V}{\alpha P_\mathrm{0}}$ of our detector to be of $\approx3.2~$kV/W ($\alpha\sim0.33$ is an attenuation factor that accounts for losses in the cryostat window, lens, and the low-pass Zitexm shield filtering thermal radiation from the room~\cite{resterahertz}). The $R_{V}$ obtained in this way provides a lower bound for the responsivity and is usually termed extrinsic, i.e., the calculations assume that the full power delivered to the device antenna is funneled into the channel. The obtained value is comparable to the commercially available NbN-based HEBs~\cite{Scontel}. In this linear regime, we have also estimated the noise equivalent power ($NEP$) of our NbSe$_2$ bolometer using a standard relation~\cite{NEP_calc} $NEP=\alpha P_\mathrm{0}/(SNR\sqrt{B})\approx~$7pW$/ \sqrt{Hz}$, where $SNR$ is the signal to noise ratio and $B$ is the measurement bandwidth. Finally, we further note in passing, that estimates based on the isothermal method suggest that only $25\%$ of the incident power is absorbed by NbSe$_2$ (\supl{Supplementary Information}). 
.

\begin{figure*}[ht!]
	\centering\includegraphics[width=0.7\linewidth]{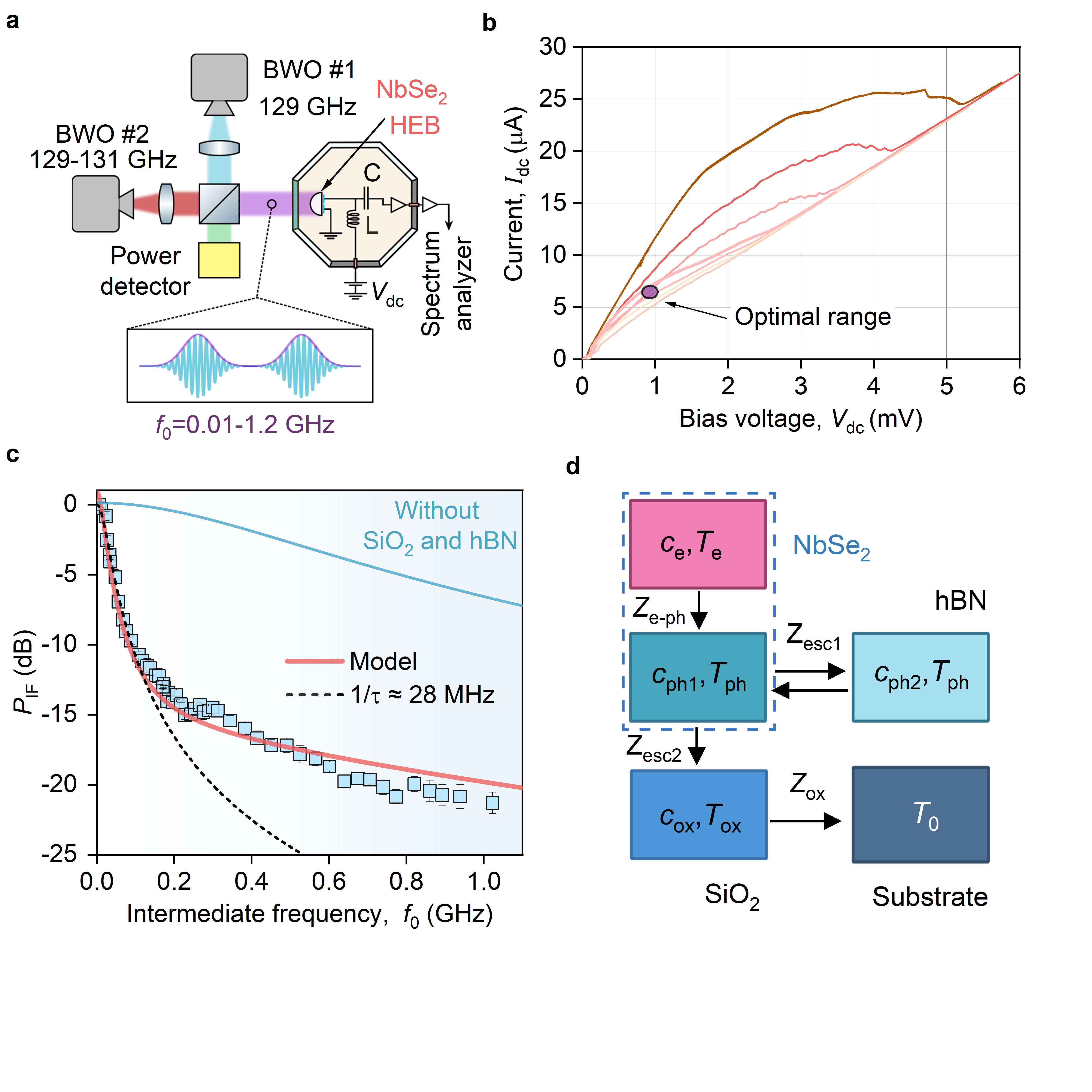}
	\caption{\textbf{Heterodyne mixing. }\textbf{a,} Typical measurement scheme for heterodyne mixing. Tunable BWOs are used as the local oscillator (LO) and signal source (RF) in the frequency range from 129 - 131 GHz. The device is located in a closed-loop cryostat and mounted on a hemispherical silicon lens. \textbf{b,} $I-V$ curves near $T_c$ under varying power values of the LO ranging from 0 to 3~$\mu$W, the shaded area shows the optimal bias voltage range for heterodyne measurement. \textbf{c,} The superconducting HEB output power vs intermediate frequency $f_0$. Pink solid line: energy relaxation model with zero fitting parameters. Black dashed line: Lorentzian decay with a characteristic 3dB roll-off frequency of 28 MHz. Solid blue line: our model's prediction of the photoresponse for the NbSe$_2$ microwire brought in direct contact to the bath, i.e., without hBN and SiO$_2$ buffer.  \textbf{d,} Thermal model for cooling path in hBN-encapsulated NbSe$_2$ devices. Electrons are thermalized via electron-phonon coupling with phonons in the NbSe$_2$ layer, which are coupled to phonons in the hBN layer. Subsequent thermalization with the bath occurs through the amorphous SiO$_2$ layer.}
	\label{Fig3}
\end{figure*}

Another crucial parameter of superconducting HEBs is their response time~\cite{Gerhenzon_electron_heat,Semenov_heb_rev,Goltsman_bol,Tretyakov_wide} $\tau$. To determine $\tau$ in our NbSe$_2$ detectors we used a standard heterodyne mixing scheme~\cite{lorentz_fit} (\supl{Supplementary Information}). In short, the scheme employs two BWOs called signal (RF) and local (LO) oscillators, which generate radiation with slightly dissimilar frequencies $f_\mathrm{RF}=130\pm1~$GHz and $f_\mathrm{LO}=129~$GHz respectively (Fig.~\ref{Fig3}a). RF and LO beams are directed onto the antenna-coupled sample, where their interference causes the beating in the absorbed power with an intermediate frequency (IF) $f_\mathrm{0}$ = $\left|f_\mathrm{RF} - f_\mathrm{LO}\right|$. By varying the frequency of the LO, the IF can be tuned from 10 MHz to 1.2 GHz, thus enabling the measurements of the frequency range within which the bolometer can respond. 

To perform the heterodyne mixing experiments, we biased our NbSe$_2$ microwires using an isolated voltage source connected to the DC port of the bias tea (Fig.~\ref{Fig3}a). Voltage bias eliminates hysteretic behavior (Fig.~\ref{Fig1}d) and enables smooth $I_\mathrm{dc}-V_\mathrm{2pt}$ curves critical for heterodyne mixing~\cite{Voltage-bias} (see Fig.~\ref{Fig3}b and \supl{Supplementary Information}). The photovoltage amplitude $V_\mathrm{IF}$ at $f_\mathrm{0}$ is read out using a spectrum analyzer connected to the RF port of the bias tea (18 GHz bandwidth) through the cryogenic amplifier. Figure~\ref{Fig3}c shows the results of such measurements for varying $f_\mathrm{0}$ obtained in the optimal operation regime of one of our NbSe$_2$ bolometers. In our experiments, we chose this optimal regime as the maximal response region (purple area in Fig.~\ref{Fig3}b). Upon increasing, $f_\mathrm{0}$, the $V_\mathrm{IF}$ decreases and becomes indistinguishable from noise above 1 GHz. Similar results were obtained for other NbSe$_2$ devices which are presented in \supl{Supplementary Information}. 

To gain insight into the processes that limit the bandwidth of our NbSe$_2$ bolometers, we conducted a standard analysis and attempted to fit the frequency dependence depicted in Fig.~\ref{Fig3}c using Lorentzian functions. Unexpectedly, the data could not be precisely fitted with a single cut-off frequency: while the low-frequency data aligned with a 28 MHz cut-off Lorentzian (dashed line), the high-frequency range notably deviated from this trend. This behavior proved consistent in both partially- and fully-encapsulated NbSe$_2$ devices and was notably absent in conventional NbN superconducting HEBs tested under the same conditions. Such a deviation from the typical Lorentzian decay response is uncommon for superconducting HEBs, suggesting special energy relaxation pathways in our NbSe$_2$ devices, which we now proceed to analyze.

In sputtered superconducting HEBs, incident THz radiation is absorbed by electrons, which is followed by electron thermalization and the subsequent release of excess energy to phonons which typically maintain thermal equilibrium with bath. The timeframe of this energy escape determines the response time of the bolometers. However, in van der Waals (vdW) heterostructures, a different scenario arises. Phonons might not achieve equilibrium with the substrate, primarily due to the vdW gap between the superconductor and the SiO$_2$ layer upon which they are typically deposited. This gap could potentially act as a barrier, delaying the energy relaxation of overheated electrons. Furthermore, within the amorphous SiO$_2$ layer thermal relaxation might also be hampered. Here, the phonon propagation tends to be diffusive, resulting from scatterings on defects, boundaries, and two-level systems typical of amorphous structures, contrasting with the faster, ballistic relaxation seen in conventional thermal bath media~\cite{Baeva_SIO2}. Considering these observations, we develop a model elucidating the observed deviation from the expected single-frequency roll-off response in our devices.

The thermal relaxation path's diagram is depicted in Figure~\ref{Fig3}d. We posit that each subsystem has a distinct temperature, governed by the quasi-equilibrium condition when the energy relaxation time scales within a subsystem are faster than the relaxation times between subsystems~\cite{Sidorova_bottleneck}. The electron subsystem in NbSe$_2$ is characterized by an electron temperature $T_e$ and a heat capacity $c_{e}$. Concurrently, the phonon subsystem in NbSe$_2$ has a phonon temperature $T_{ph}$ and a heat capacity $c_{ph1}$. The electron and phonon subsystems in NbSe$2$ interact via the electron-phonon coupling, represented by the thermal resistance $Z_{e-ph}$. The phonon subsystem in the hBN layer, characterized by a phonon temperature $T_{ph}$ and heat capacity $c_{ph2}$, is connected to the phonons in NbSe$2$ through the thermal boundary resistance $Z_{esc1}$. Both these phonon subsystems interface with the substrate through the thermal boundary resistance $Z_{esc2}$. The phonon subsystem in the SiO$_2$ layer, characterized by the average phonon temperature $T_{ox}$ and a heat capacity $c_{ox}$, influences the thermal relaxation via the thermal resistance $Z_{ox}$. The crystalline Si section of the substrate is regarded as a thermal reservoir at a temperature $T_0$. Consequently, the bolometer is interpretable as a distributed system, and its behavior can be modeled using the thermal analog of Ohm's law: $\delta T = P_{abs} Z_{th}$, where $\delta T$ represents the electronic temperature deviation from the bath, $P_{abs}$ is the absorbed power, and $Z_{th}$ signifies the effective lumped-element thermal impedance. $\delta T$ leads to the bolometer voltage response $V_{IF}$, which equates to the IF output power $P_{IF}$ at the load resistance $R_L$ of the IF amplifier, given by $P_{IF} = V_{IF}^2/(2R_L)$. The electrical analogy of the thermal circuit under discussion can be found in the \supl{Supplementary Information}.

To model the response spectrum, in \supl{Supplementary Information} we provided the full determination of the effective thermal impedance $Z_{th}$. The comparison of the model and the experimental data is shown in Fig.~\ref{Fig3}c that plots $P_{IF}$ as a function of $f_\mathrm{0}$ using the frequency dependence of $Z_{th}$ (pink line). The model shows a good agreement with the experiment considering that it does not involve any fitting parameters (all the parameters are taken from independent transport measurements provided in \supl{Supplementary Information} and literature). This result can be interpreted in the framework of the discussed thermal model as follows. The main factor limiting the bandwidth of our NbSe$_2$ bolometers is related to the larger contribution of the thermal resistance of the SiO$_2$ layer compared to other relaxation processes ($Z_{e-ph}, Z_{esc2} < Z_{ox}$), which may be due to both the low thermal conductivity and relatively large thickness of the SiO$_2$ layer (280 nm). At higher frequencies, the observed response also points to other factors limiting the bandwidth in NbSe$_2$ bolometers, in particular, the heating of hBN phonons and their participation in the overall energy relaxation in the device. In our model, we also obtain relatively fast electron-phonon relaxation and phonon escape to the substrate, for which the following characteristic time scales were obtained: $\tau_{e-ph}  \approx 89$\,ps and $\tau_{esc}\approx 60$\,ps (details given in \supl{Supplementary Information}). Importantly, the effective relaxation time due to the e-ph scattering and the phonon escape to the oxide is $\tau = \tau_{e-ph} + \tau_{esc} c_e/c_{ph}\approx0.28$\,ns (here $c_e/c_{ph} = 3.2$), which corresponds to much higher values of the cut-off frequency ($568$\,MHz) which is close  to the our model's prediction of the photoresponse for the NbSe$_2$ microwire brought in direct contact to the bath (blue line in Fig.~\ref{Fig3}c), that we envision can be achieved if the devices are assembled on substrates with larger thermal conductance. 
 
In conclusion, we have demonstrated the use of NbSe$_2$ vdW superconductors in the construction of THz bolometers. By improving the interface between NbSe$_2$ and metal antenna sleeves, we mitigated the impedance mismatch between this vdW superconductor and the RF readout circuitry, thus enabling us to attain strong photoresponse at 0.13 and 2.5 THz. Moreover, this RF readout capability facilitated precise measurements of our device's response time, revealing the deviation from the expected for superconducting HEBs single-frequency roll-off response. This deviation is attributed to a slow pathway for heat dissipation from NbSe$_2$ to the bath through a relatively thick SiO$_2$ layer on which our devices are assembled. Our work opens up the family of vdW superconductors to supoerconducting HEB technology and highlights NbSe$_2$ as a promising platform for THz applications. It would also be interesting to conduct similar experiments with atomically thin high-T$_c$ vdW superconductors, which have recently gained momentum in the field of infrared optoelectronics~\cite{Merino_2023,Charaev2023}.

\vspace{1em}

\noindent\rule{6cm}{0.4pt}

*Correspondence to: dab@nus.edu.sg.

\section*{Acknowledgements}
% The authors thank Maria Sidorova and Alessandro Principi for productive discussions. 

\section*{Data availability}
All data supporting this study and its findings are available within the article and its Supplementary Information or from the corresponding authors upon reasonable request.

\section*{Author contributions}

\section*{Competing interests}
The authors declare no competing interests.

\bibliography{Bibliography_text}

\end{document}